\documentclass[useAMS,usenatbib]{mn2e}

\usepackage{times}
\usepackage{graphicx}
\usepackage{subfigure}
\usepackage{amsmath}

\newcommand{\gsim}{\mathrel{\hbox{\rlap{\lower.55ex \hbox {$\sim$}}
                   \kern-.3em \raise.4ex \hbox{$>$}}}}
\newcommand{\lsim}{\mathrel{\hbox{\rlap{\lower.55ex \hbox {$\sim$}}
                   \kern-.3em \raise.4ex \hbox{$<$}}}}

\title[Dust dynamics during protostellar collapse]{On the dynamics of dust during protostellar collapse}
\author[M. R. Bate \& P. Lor\'en-Aguilar]{Matthew R. Bate$^{1}$\thanks{E-mail:
mbate@astro.ex.ac.uk (MRB);  pablo@astro.ex.ac.uk (PLA)} and Pablo Lor\'en-Aguilar$^{1}$\\
$^{1}$ School of Physics and Astronomy, University of Exeter, Stocker
Road, Exeter EX4 4QL  
}

\date{Accepted by MNRAS}
\begin{document}
\maketitle
\begin{abstract}
The dynamics of dust and gas can be quite different from each other when the dust is poorly coupled to the gas.  In protoplanetary discs, it is well known that this decoupling of the dust and gas can lead to diverse spatial structures and dust-to-gas ratios.  In this paper, we study the dynamics of dust and gas during the earlier phase of protostellar collapse, before a protoplanetary disc is formed.  We find that for dust grains with sizes $\lsim 10 ~\mu$m, the dust is well coupled during the collapse of a rotating, pre-stellar core and there is little variation of the dust-to-gas ratio during the collapse.  However, if larger grains are present, they may have trajectories that are very different from the gas during the collapse, leading to mid-plane settling and/or oscillations of the dust grains through the mid-plane.  This may produce variations in the dust-to-gas ratio and very different distributions of large and small dust grains at the very earliest stages of star formation, if large grains are present in pre-stellar cores.
\end{abstract}

\begin{figure*}
\centering \vspace{0cm} \vspace{-0.5cm}
    \includegraphics[width=8cm]{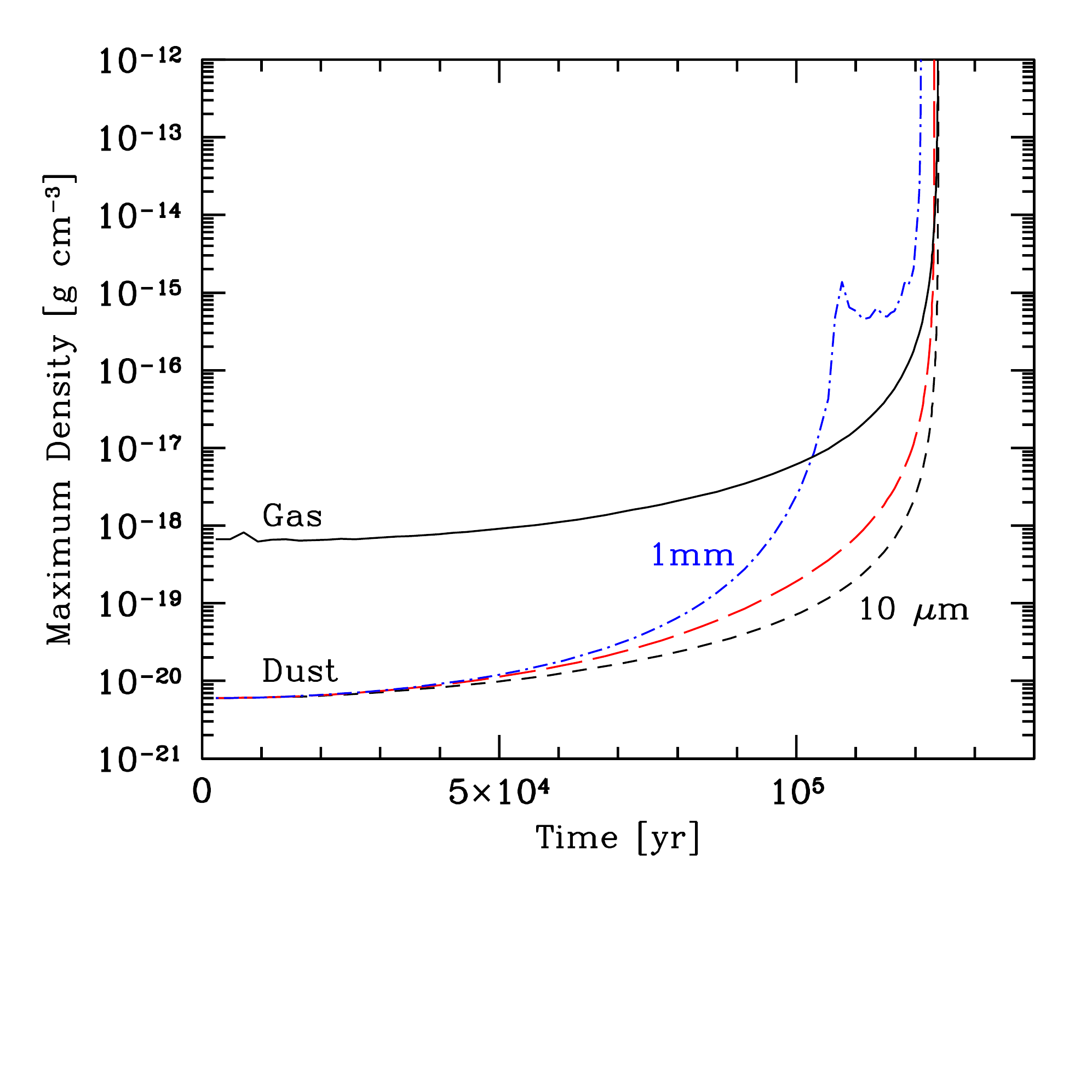} \vspace{0cm}
   \includegraphics[width=8cm]{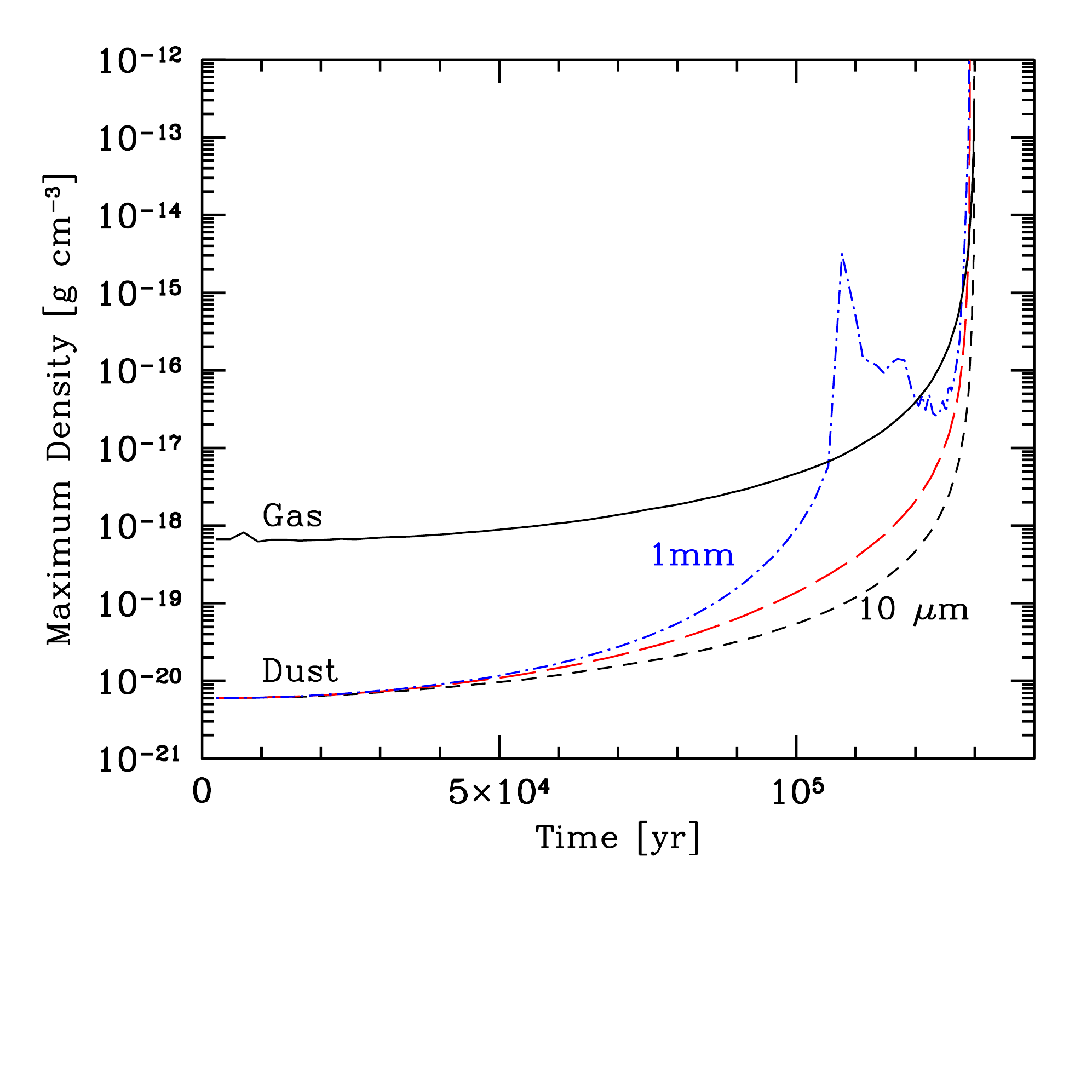} \vspace{-1.7cm}
\caption{The time evolution of the maximum density of the gas and dust.  The left panel shows the results from the slowly rotating calculations ($\beta=0.02$), while the right panel shows the more rapidly rotating case ($\beta=0.08$).  The solid black line gives the maximum gas density (from the $s=10~\mu$m calculations), while the other lines give the maximum dust density when using grain sizes of 10~$\mu$m (short-dashed, black lines), 100~$\mu$m (long-dashed, red lines), and 1~mm (dot-dashed, blue lines).  Larger grains are more poorly coupled to the gas and, therefore, collapse more quickly.  The maximum density of the 1~mm grains is non-monotonic because the dust passes through, and oscillates around, the mid-plane. We only show the gas density from the $s=10~\mu$m calculation, since its evolution does not differ substantially between calculations with different dust types.}
\label{fig:density_time}
\end{figure*}

\section{Introduction}
\label{introduction}

Dust plays important roles in both star formation and planet formation.  In the former case, dust thermal emission dominates molecular gas cooling at number densities $n \gsim 10^4$~cm$^{-3}$ \citep[e.g.][]{Goldsmith2001}.  In the later case, it is believed that the growth of interstellar grains in the comparatively high-density environments provided by protoplanetary discs eventually leads to planet formation.

The dynamical interaction between gas and dust is manifest primarily as a drag force.  The drag force experienced by dust grains moving through a gas depends mainly on the size of the dust grains and the density of the gas \citep{Whipple1972, Weidenschilling1977}.  If the mean free path of the gas molecules is larger than the dust particle's radius, the characteristic timescale for the decay of the dust particle's speed relative to the gas, the stopping time, can be expressed as
\begin{equation}
t_{\rm s} = \frac{\hat{\rho}_{\rm s} s}{\rho_{\rm G} v_{\rm th} },
\label{eq:stoppingtime}
\end{equation}
where, for simplicity, we have assumed that the dust-to-gas ratio is small, $s$ is the radius of the dust particle, $\hat{\rho}_{\rm s}$ is the intrinsic density of the dust particle, $\rho_{\rm G}$ is the gas density, and the velocity of the gas molecules due to thermal motion is
\begin{equation}
v_{\rm th} = \sqrt{\frac{8 k_{\rm B} T}{\pi \mu m_{\rm H}}},
\end{equation}
where $T$ is the gas temperature, $\mu$ is the mean molecular weight of the gas, $m_{\rm H}$ is the atomic mass of hydrogen, and $k_{\rm B}$ is Boltzmann's constant. Thus, small grains have short stopping times and are better coupled to the gas dynamically, while larger grains have longer stopping times and may be only weakly affected by the drag force.  

Usually, it is assumed that the dust in the interstellar medium (ISM) consists of small grains ($\sim 0.1 ~\mu$m) so that in molecular clouds the dust is both well-mixed with, and well-coupled to, the gas.  For example, using equation \ref{eq:stoppingtime}, the stopping time of such grains at densities of $n = 10^4$~cm$^{-3}$ is $t_{\rm s}\approx 1000$~yrs, which is expected to be much shorter than the dynamical timescale on which the gas density changes substantially (e.g. the free-fall time is $\approx 3 \times 10^5$ yrs), except in shocks.  Recently, however, \cite{Hopkins2014} and \cite{HopLee2016} pointed out that in the low-density ISM even small grains can become poorly coupled (e.g.\ $n = 1$~cm$^{-3}$; $t_{\rm s}\approx 1 \times 10^7$~yrs),  and, therefore, variations in the dust-to-gas ratio may be expected.

In protoplanetary discs, the high densities are thought to allow substantial growth of dust grains \citep{WeiCuz1993}.  Once a population consisting of different grain sizes has developed, because grains of different sizes are more or less coupled to the gas, the dynamics of the dust, and therefore the distribution of the dust, is expected to depend on the grain size.  Since grain growth is thought to occur more rapidly at higher densities, and dust migrates in discs toward locations of higher pressure, it may be expected that larger grains are found towards the inner parts of protoplanetary discs \citep[e.g][]{Birnstieletal2010}.  Observations provide evidence for radial variations of both the typical dust size \citep[e.g.][]{Tazzarietal2016} and the dust-to-gas ratio \citep[e.g.][]{Andrewsetal2012}. Other predicted effects of the different dynamics of dust and gas include dust trapping in vortices \citep[e.g.][]{BarSom1995}, in spiral density waves \citep[e.g.][]{Riceetal2004}, and in rings near planets \citep[e.g.][]{PaaMel2004}.

Although it is generally accepted that dust in the interstellar medium has a typical maximum size of $s \sim 0.1~\mu$m \citep*[e.g.][]{MatRumNor1977}, some observations of molecular clouds and young protostellar objects provide evidence for a population of larger grains \citep[e.g.][]{LehMat1996}.  Recent infrared observations of `cloudshine' from molecular clouds \citep{FosGoo2006} indicate that dust grains with sizes $> 1 ~\mu$m may present in molecular clouds \citep{Paganietal2010, Steinackeretal2014, Steinackeretal2015, Andersenetal2013}, although recent work has shown that these observations may be explained by the accretion of hydrogen-rich carbon mantles without the necessity of appealing to large grains \citep{Ysardetal2016}.  Observations of mid-infrared extinction in both the ISM and dense clouds also suggest the presence of grains with sizes $> 1 ~\mu$m \citep*{WanLiJia2015a,WanLiJia2015b}.   Some dense molecular cores and filaments have low observed values of the emissivity spectral index, $\beta$, at (sub-)mm wavelengths which may due to the presence of mm-size dust grains \citep{Miettinenetal2012, Schneeetal2014}.  In Class 0 and I protostellar objects, despite their young age ($\sim 10^5$ yrs), their envelopes apparently contain mm-size dust grains (\citealt{Kwonetal2009}; \citealt*{ChiLooTob2012}; \citealt{Miotelloetal2014}), despite the difficulties of explaining this with theoretical models of grain growth \citep*[e.g.][]{WonHirLi2016}.  Theoretically, large dust grains can grow in molecular clouds, but only if they are both dense and long-lived \citep[e.g.][]{Ormeletal2009}.

If some large grains are present in pre-stellar cores, then they may be coupled weakly enough to the gas to undergo significantly different dynamical evolution during protostellar collapse compared to the gas.  For example, taking a pre-stellar core with density $n = 10^5$~cm$^{-3}$, the stopping time of $100 ~\mu$m grains is  $t_{\rm s} \approx 10^5$~yrs, which is equal to the free-fall time.

In this paper, we consider the dynamics of dust grains during the initial collapse of a gravitationally unstable molecular cloud core to determine how the distribution of dust grains that are initially uniformly mixed with the gas varies with the size of the dust grains.  As expected from the above analytical estimate, we find that dust with sizes $s \gsim 100 ~\mu$m, if present, evolves differently from the gas and its distribution may be quite dissimilar from that of the gas even before stellar core formation occurs.  In Section 2, we describe our numerical method and the initial conditions for our calculations.  Our results are presented in Section 3, and in Section 4 we draw our conclusions.

\section{Method}
\label{sec:method}

The calculations presented here were performed 
using a three-dimensional smoothed particle
hydrodynamics (SPH) \citep{Lucy1977, GinMon1977} code based on the original 
version of \citeauthor{Benz1990} 
(\citeyear{Benz1990}; \citealt{Benzetal1990}), but substantially
modified as described in \citet*{BatBonPri1995},
\cite{PriBat2007}, and 
parallelised using both OpenMP and MPI.

Gravitational forces between particles and a particle's 
nearest neighbours are calculated using a binary tree.  
The smoothing lengths of particles are variable in 
time and space, set iteratively such that the smoothing
length of each particle 
$h = 1.2 (m/\rho)^{1/3}$ where $m$ and $\rho$ are the 
SPH particle's mass and density, respectively
\citep{PriMon2007}.  To reduce numerical shear viscosity, we use the
\cite{MorMon1997} artificial viscosity
with $\alpha_{\rm_v}$ varying between 0.1 and 1 while $\beta_{\rm v}=2 \alpha_{\rm v}$
\citep[see also][]{PriMon2005}.

The code can evolve both gas and dust dynamically using two populations
of SPH particles.  Integration of the drag force between the gas
and the dust can be carried out either explicitly  \citep[see][]{Ayliffeetal2012}
using the method of \cite{LaiPri2012a}, or
using the semi-implicit method of \cite{LorBat2014, LorBat2015b}.
In this paper, due to the fact that we are consideing relatively small grains,
we use the semi-implicit integration method.
The SPH equations are 
integrated using a second-order Runge-Kutta-Fehlberg 
integrator \citep{Fehlberg1969} with global time steps, as the
semi-implicit method cannot yet use individual time steps.

\begin{figure*}
\centering \vspace{-0.25cm} \vspace{0cm}
    \includegraphics[height=9.5cm]{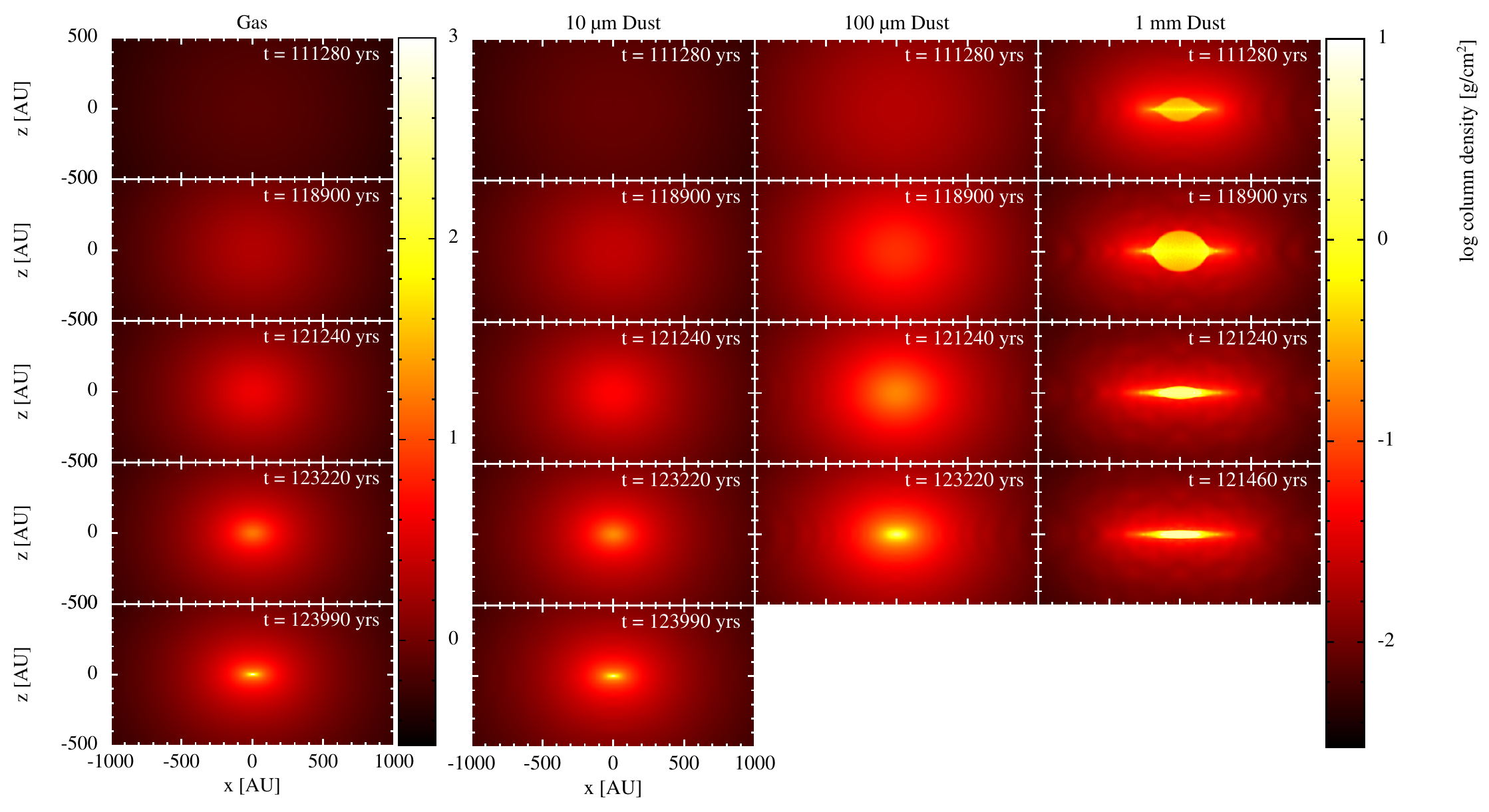} \vspace{-0.25cm}
\caption{The evolution of the column density (parallel to the rotation axis) of the gas and dust for the calculations with slow initial rotation rates ($\beta=0.02$).  From left to right, the columns display the gas column density (from the $s=10~\mu$m calculation), and the dust column densities when using grain sizes of $s=10, 100~\mu$m, and 1~mm at various times.  The colour scales for the gas and the dust have been chosen so that for a constant dust-to-gas ratio of 0.01 the colours will be identical.  When the colours are brighter in the dust image than the associated gas image, the dust-to-gas ratio has increased.  With the smallest grains ($s=10~\mu$m), the dust and gas are well-coupled and the dust column density closely follows that of the gas.  Grains with intermediate sizes  ($s=100~\mu$m) collapse somewhat more quickly than the gas, enhancing the central dust-to-gas ratio.  The largest grains  ($s=1~$mm) collapse even more quickly, and are so poorly coupled to the gas that they perform vertical oscillations through the mid-plane before settling into a large dust disc and becoming much more concentrated than the gas.  We only show the gas from the $s=10~\mu$m calculation, since its distribution does not differ substantially between the calculations.}
\label{fig:image002}
\end{figure*}

\begin{figure*}
\centering \vspace{-0.25cm} \vspace{0cm}
    \includegraphics[height=9.5cm]{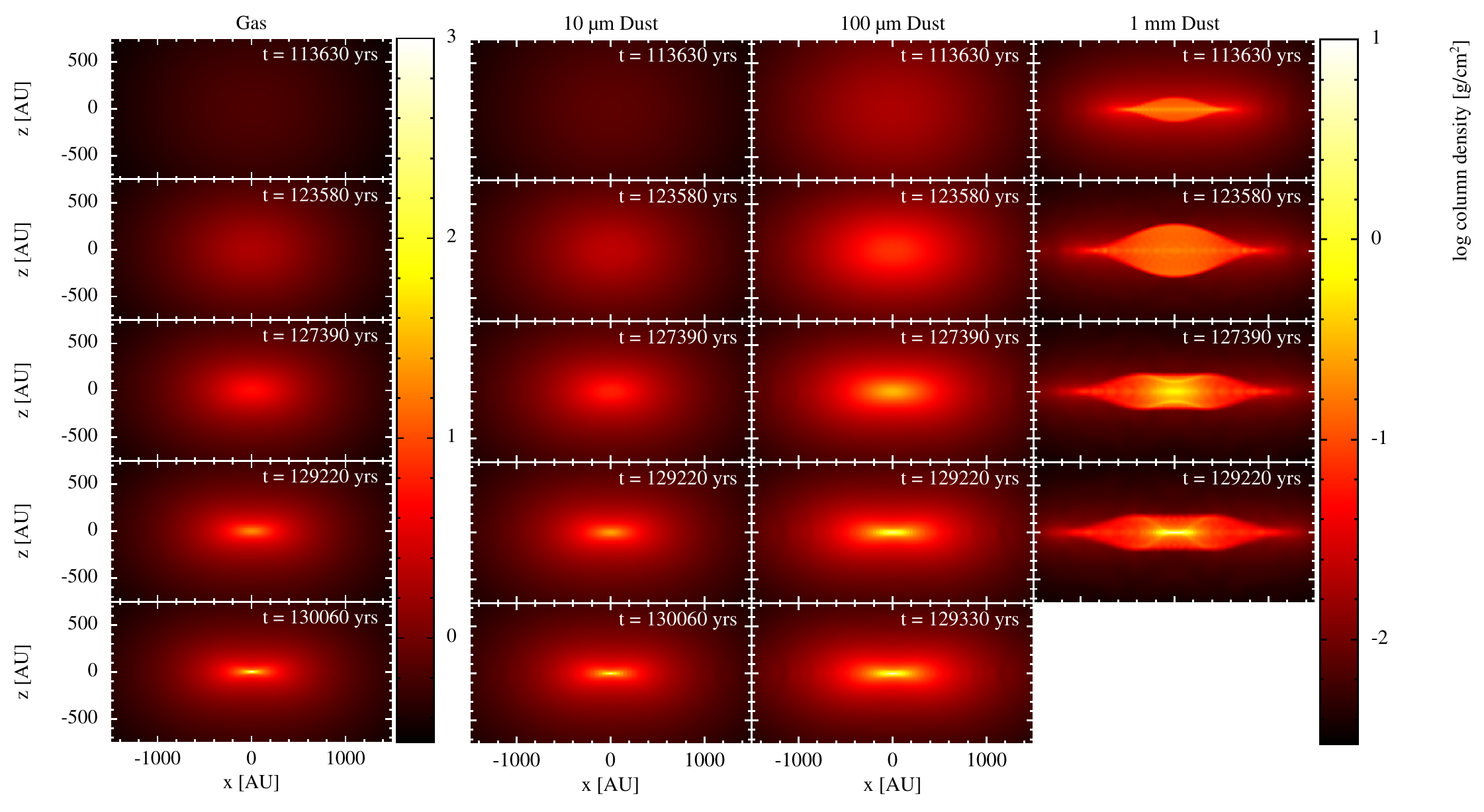} \vspace{-0.25cm}
\caption{The evolution of the column density (parallel to the rotation axis) of the gas and dust for the calculations with fast initial rotation rates ($\beta=0.08$).  From left to right, the columns display the gas column density (from the $s=10~\mu$m calculation), and the dust column densities when using grain sizes of $s=10, 100~\mu$m, and 1~mm at various times.  The colour scales for the gas and the dust have been chosen so that for a constant dust-to-gas ratio of 0.01 the colours will be identical.  When the colours are brighter in the dust image than the associated gas image, the dust-to-gas ratio has increased.  With the smallest grains ($s=10~\mu$m), the dust and gas are well-coupled and the dust column density closely follows that of the gas.  Intermediate-sized grains  ($s=100~\mu$m) collapse more quickly than the gas, producing a disc-like distribution earlier than the gas, with a somewhat larger radius.  The largest grains  ($s=1~$mm) collapse even more quickly, and are so poorly coupled to the gas that they perform vertical oscillations through the mid-plane before settling into a disc which is much larger than the gas disc.   We only show the gas from the $s=10~\mu$m calculation, since its distribution does not differ substantially between the calculations.}
\label{fig:image008}
\end{figure*}

\subsection{Initial conditions, equation of state, and resolution}
\label{initialconditions}

Our initial conditions for the molecular cloud core consist of unstable Bonnor-Ebert spheres. 
We choose 5-M$_\odot$ clouds with an inner to outer density contrast of 20 and a radius of 0.1~pc \citep[the same initial conditions as used by][]{BatKet2015}.  The clouds are contained by spherical, reflective boundary conditions (modelled using ghost particles).  The clouds are set in uniform rotation initially, and we perform calculations with two different rotation rates of $5.06 \times 10^{-14}$ and $1.012\times 10^{-13}$~rad~s$^{-1}$.  These correspond to ratios of rotational to gravitational potential energy with magnitudes of $\beta=0.02$ and $0.08$, respectively.

The thermal evolution of the gas was modelled using a barotropic equation of state, where the gas pressure was given by
\begin{equation}
P = \left\{ \begin{array}{l l} c_\text{s,0}^2\rho; 		    &  \rho < \rho_\text{c}, \\
                                           c_\text{s,0}^2\rho_\text{c}\left(\rho             /\rho_\text{c}\right)^{7/5};      &  \rho_\text{c} \leq \rho,\end{array}\right.
\end{equation}
where $c_\text{s,0} = 1.87\times 10^4$~cm~s$^{-1}$ is the initial isothermal sound speed of gas with a temperature of 10~K (the mean molecular weight is $\mu=2.38$), and $\rho_\text{c} = 10^{-13}$ g~cm$^{-3}$. This equation of state is designed to mimic the evolution of the pressure in collapsing molecular clouds \citep*{Larson1969,MasInu2000}.   In fact, our calculations only follow the collapse until shortly after the density exceeds $\rho_{\rm c}$, so the gas is isothermal at 10~K throughout the calculations, except right at the end of the calculations when the temperature at the centre of the clouds begins to rise.  The maximum temperature reached in any of the calculations is $\approx 60$~K.

The calculations whose results are presented in this paper employed $1 \times 10^7$ SPH gas particles and $3\times 10^5$ SPH dust particles to model the cloud.  Our gas resolution is two orders of magnitude larger than the number or particles required to resolve the local Jeans mass throughout the calculation (\citealt{BatBur1997, Trueloveetal1997, Whitworth1998}; \citealt*{HubGooWhi2006}).  We also performed lower resolution calculations using $3 \times 10^6$ and $1 \times 10^5$ particles, respectively, and found no significant change in the results with different resolutions.  

The Bonnor-Ebert density distribution of the gas particles is set up using equal-mass particles which are placed on a uniform cubic lattice that is deformed radially to achieve the required density profile.  The dust particles are distributed in the same manner, but dust particles are excluded from the outer 10\% of the cloud by radius to avoid any potential problems with boundaries.  In fact, this may be closer to reality than having a uniform dust-to-gas ratio in the outer parts of the cloud, since \cite{WhiBat2002} have shown that radiation pressure from the interstellar radiation field may be expected to push dust particles from the low-density, outer regions of a molecular cloud into the cloud.  This produces a deficit of dust in the outer parts, and a sharper edge to the dust distribution than is found in the gas.  

For each of the two initial rotation rates, we perform calculations for three different intrinsic types of dust particle.  We assume spherical grains with an intrinsic density of 3.0~g~cm$^{-3}$, with radii: $s = 10, 100 ~\mu$m and 1~mm (i.e. each calculation has a single population of dust).  We assume that the Epstein drag force is valid for all our calculations.  This requires that $s < (9/4) \lambda$, where $\lambda$ is the gas mean free path, and that the relative velocity between the gas and the dust is much less than the mean thermal velocity of the gas, i.e. $|  \mbox{\boldmath{$v$}}_{\rm D} - \mbox{\boldmath{$v$}}_{\rm G} | \ll v_{\rm th}$ \citep{Weidenschilling1977}.  The former is easily satisfied throughout all of the calculations considered in this paper.  In the calculations with 100$~\mu$m grains, the maximum difference between the velocity of the dust and that of the gas at a given location can reach $\approx v_{\rm th}/2$.  In the calculations with 1~mm grains, the maximum velocity difference approaches $v_{\rm th}$.  \cite{DraSal1979} provide an expression for the drag force which is applicable for all $|  \mbox{\boldmath{$v$}}_{\rm D} - \mbox{\boldmath{$v$}}_{\rm G} |$ and accurate to within 1\%.  The effect \cite[see][]{HopLee2016} is that equation \ref{eq:stoppingtime} is multiplied by the factor
\begin{equation}
\left( 1 + \left| \frac{3}{\sqrt{8}} \frac{ \mbox{\boldmath{$v$}}_{\rm D} - \mbox{\boldmath{$v$}}_{\rm G}}{v_{\rm th}}\right|^2 \right)^{-1/2}.
\end{equation}
Therefore, at most, our calculations under-estimate the drag by up to $30$\%, but since we investigate the evolution of dust grains whose sizes (and, therefore, stopping times) differ by two orders of magnitude this small difference will not substantially alter our results. 

Within the bulk of the cloud, the dust-to-gas ratio is initially constant and set to 1/100.  Both gas and dust SPH particles contribute to the gravitational force, so the initial gas mass is 4.95~M$_\odot$, while the initial dust mass is 0.05~M$_\odot$.  With the numbers of particles given above, this means that the dust SPH particles are about a factor of three less massive than those of the gas.  We use substantially fewer dust than gas particles so that the dust smoothing lengths are initially substantially larger than the gas smoothing lengths, since with two-fluid dust/gas simulations, if the dust resolution becomes significantly smaller than the gas resolution, this can lead to artificial clumping of the dust \citep{Ayliffeetal2012}.  Only if the dust particles migrate relative to the gas can the dust resolution length potentially become smaller than that of the gas; this process does eventually result in the termination of our calculations after the formation of the first hydrostatic core  \citep{Larson1969}.

Although the calculations include the back-reaction of the dust on the gas, its effect is generally small because of low initial dust-to-gas ratio and the short dynamical time of the calculations (essentially one free-fall time).  Towards the end of the calculations with larger grain sizes the dust-to-gas ratio becomes large in the centre of the collapsing cloud and the gas density is slightly increased when compared to calculations with the smallest dust grains at the same time because of the extra mass.  However, the effect is small -- we have performed the entire suite of calculations again with an initial dust-to-gas ratio of 1/1000 and the relative distribution of dust is almost identical to that obtained with the higher dust-to-gas ratio.  Therefore, although we have performed individual calculations for each dust particle size, the resulting dust distributions should be almost identical to those that would be obtained by using a continuous size distribution of dust particles and examining the distributions dust particles with equivalent small ranges in size.

\section{Results}
\label{results}

Since the initial clouds are gravitationally unstable, the gas and dust begin to collapse.  The evolution of the maximum gas and dust densities with time are shown in Fig.~\ref{fig:density_time}.  The more rapidly rotating clouds take slightly longer to collapse, as expected.   The smallest dust grains ($s \lsim 10~\mu$m) closely follow the gas, so the maximum dust density remains a factor of 100 lower than the gas density throughout the collapse to form a first hydrostatic core.  With larger dust grains, the dust collapses more quickly than the gas, producing higher dust-to-gas ratios in the inner parts of the cloud as the collapse proceeds.  With $100~\mu$m grains, the dust-to-gas ratio exceeds 0.1 near the centre of the collapsing cloud, while with $s=1$~mm, the dust density can exceed the gas density in local regions.  Of course, for a real pre-stellar core only a small fraction of the total dust mass is expected to be in mm-size grains, if any.

In Figs.~\ref{fig:image002}  and \ref{fig:image008} we display snapshots of the evolution of the dust and gas column density in planes parallel to the rotation axis.  Animations are provided in the Supporting Information that accompanies the paper.  The smallest grains ($s \lsim 10~\mu$m) closely follow the gas, but the larger grains develop different spatial distributions.  The $100~\mu$m grains are coupled well enough to the gas that they collapse monotonically toward the mid-plane, but they do so more quickly than the gas, resulting in a dusty disc-like structure that has an increased dust-to-gas ratio (particularly with the higher initial rotation rate).  The 1~mm grains are so poorly coupled to the gas that they initially pass through the mid-plane and perform damped vertical oscillations through the mid-plane while they slowly settle.  This produces increases in the local dust-to-gas ratio by factors of up to $10^5$ times the initial value in small regions where the dust converges.  These extreme values are confined to short periods as the dust passes through the mid-plane (see Fig.~\ref{fig:density_time}), but increases in the initial dust-to-gas ratio by factors of $100-1000$ are typical in a large volume in the vicinity of the mid-plane within $500-1000$~AU of the centre of the cloud  (depending on the initial rotation rate; Figs.~\ref{fig:image002} and \ref{fig:image008}).  

Again it is important to emphasise that for simplicity we have taken the initial dust-to-gas ratio to be 1/100 for each of our individual calculations.  In reality, it is likely that most of the total dust mass will be in small grains, with only a fraction in large grains.  If this is the case, then even if the ratio of the density of large grains to the gas density increases by several orders of magnitude, the effect on the {\it total} dust-to-gas ratio may be much smaller.  Observationally, measuring dust-to-gas ratios is extremely difficult as it relies on accurate chemical and radiative transfer modelling, but there are many examples of protoplanetary discs which may have substantially enhanced dust-to-gas ratios \citep[e.g.][and references therein]{WilBes2014, Bonebergetal2016}.

\section{Conclusions}
\label{conclusions}

We have shown that if large grains $s \gsim 100~\mu$m are present in pre-stellar cores, their dynamical evolution will be very different to that of the gas and of smaller dust grains as the core collapses to form a star, even at the earliest stages of star formation.  If large grains are present in significant numbers, this would result in a dust-to-gas ratio greater than the initial value in the inner parts of the molecular cloud core even before the first hydrostatic core forms.  It also means that right from the earliest phases of star formation, grains of different sizes could be spatially segregated, even in the absence of significant grain growth.  Large grains would be more abundant relative to small grains near the centre of the collapsing molecular cloud core than they are in the outer parts of the cloud.  They will also tend to be distributed in a more flattened disc-like structure than either the gas or the small grains.

The dataset consisting of the output and analysis files from the calculations presented in this paper have been placed in the University of Exeter's Open Research Exeter (ORE) repository.

\section*{Acknowledgments}

This work was supported by the European Research Council under the European Community's Seventh Framework Programme (FP7/2007-2013 Grant Agreement No. 339248).  This work used the DiRAC Complexity system, operated by the University of Leicester, which forms part of the STFC DiRAC HPC Facility (www.dirac.ac.uk). This equipment is funded by BIS National E-Infrastructure capital grant ST/K000373/1 and  STFC DiRAC Operations grant ST/K0003259/1. DiRAC is part of the National E-Infrastructure. This work also used the University of Exeter Supercomputer, a DiRAC Facility jointly funded by STFC, the Large Facilities Capital Fund of BIS and the University of Exeter.

\bibliography{mbate}

\end{document}